\documentclass[conference]{IEEEtran}

\IEEEoverridecommandlockouts
\usepackage{cite}
\usepackage{amsmath,amssymb,amsfonts}
\usepackage{algorithmic}
\usepackage{graphicx}
\usepackage{textcomp}
\usepackage{xcolor}
\def\BibTeX{{\rm B\kern-.05em{\sc i\kern-.025em b}\kern-.08em
    T\kern-.1667em\lower.7ex\hbox{E}\kern-.125emX}}

\usepackage{cleveref}
\crefname{figure}{Fig.}{Figs.}
\crefname{section}{Section}{Sections}

\usepackage[normalem]{ulem}
\usepackage{csquotes}
\usepackage{ulem}

\usepackage{multirow}

\usepackage{subcaption}

\usepackage[nolist]{acronym}

\newcommand{\argmin}[1]{\underset{#1}{\operatorname{arg}\,\operatorname{min}}\;}

\begin{document}
% Define acronyms
\begin{acronym}
    \acro{rl}[RL]{Reinforcement Learning}
    \acro{ppo}[PPO]{Proximal Policy Optimization}
    \acro{drl}[DRL]{Deep Reinforcement Learning}
    \acro{mpc}[MPC]{Model Predictive Control}
    \acro{dof}[DOF]{Degree of Freedom}
    \acro{lqr}[LQR]{Linear-quadratic regulator}
    \acro{lqi}[LQI]{Linear-quadratic-integral regulator}
    \acro{mdp}[MDP]{Markov Decision Process}
    \acro{sb3}[SB3]{Stable Baselines3}
\end{acronym}

\title{Comparison of Model Predictive Control and Proximal Policy Optimization for a 1-DOF Helicopter System
%\\{\footnotesize \textsuperscript{*}Note: Sub-titles are not captured in Xplore and should not be used}
%\thanks{Supported by the Christian Doppler Research Association (JRC ISIA). This preprint has not undergone peer review or any post-submission improvements or corrections.}
}

\author{
\IEEEauthorblockN{Georg Schäfer\IEEEauthorrefmark{1}\IEEEauthorrefmark{2}, Jakob Rehrl\IEEEauthorrefmark{1}, Stefan Huber\IEEEauthorrefmark{1}, Simon Hirlaender\IEEEauthorrefmark{2}}
\IEEEauthorblockA{\IEEEauthorrefmark{1}Josef Ressel Centre for Intelligent and Secure Industrial Automation, Salzburg, Austria}
\IEEEauthorblockA{\IEEEauthorrefmark{2}Paris Lodron University of Salzburg, Salzburg, Austria \\
georg.schaefer@fh-salzburg.ac.at}
}

\maketitle

\begin{abstract}
This study conducts a comparative analysis of Model Predictive Control (MPC) and Proximal Policy Optimization (PPO), a Deep Reinforcement Learning (DRL) algorithm, applied to a 1-Degree of Freedom (DOF) Quanser Aero 2 system. Classical control techniques such as MPC and Linear Quadratic Regulator (LQR) are widely used due to their theoretical foundation and practical effectiveness. However, with advancements in computational techniques and machine learning, DRL approaches like PPO have gained traction in solving optimal control problems through environment interaction. This paper systematically evaluates the dynamic response characteristics of PPO and MPC, comparing their performance, computational resource consumption, and implementation complexity. Experimental results show that while LQR achieves the best steady-state accuracy, PPO excels in rise-time and adaptability, making it a promising approach for applications requiring rapid response and adaptability. Additionally, we have established a baseline for future RL-related research on this specific testbed. We also discuss the strengths and limitations of each control strategy, providing recommendations for selecting appropriate controllers for real-world scenarios.
\end{abstract}


\section{Introduction}

Control theory plays a central role in numerous engineering applications~\cite{ivanov2018survey}, enabling the design of systems to achieve a desired behavior by affecting the system dynamics. With the rapid development in computational techniques and machine learning algorithms, there is a growing interest in utilizing these advancements for control tasks. \ac{rl}, in particular, has gained interest in solving optimal control problems solely through interaction with the environment~\cite{kiumarsi2017optimal}.

In parallel, classical control techniques such as \ac{mpc} and \ac{lqr} continue to be widely used in various industrial settings due to their well-established theoretical foundation and practical effectiveness. \ac{mpc}, in particular, is favored due to its capability of systematically handling input-, output- and state-constraints, as well as measurable disturbances. Additionally, the tuning of the controller is intuitive and it is applicable to multi-input multi-output systems~\cite{maaruf2022survey, gorges2017relations}. \ac{lqr} provides a simple control law that is derived from optimality considerations utilizing linear system theory~\cite{Franklin1998}.

The objective of this study is to conduct a comparative analysis of \ac{ppo}~\cite{schulman2017proximal}, a \ac{drl} algorithm, and the classic approaches mentioned above, in the context of controlling the Quanser Aero~2 system in a 1-\ac{dof} configuration (see \cref{sec:experimental_setup}). To provide recommendations on the selection of controller, we compare these techniques in two aspects: first, dynamic response characteristics typically used in control theory are evaluated systematically. Second, less quantitative measures, such as development effort, computational resource consumption, and performance in real-world scenarios are discussed. Strengths and weaknesses of the considered \ac{drl} approach compared to the classic control methods are provided.

\subsection{Related Work}

Several control strategies have been proposed in the literature for the Quanser Aero~2 system. Subramanian et al. introduced a controller based on \ac{lqr} for robust control of the system~\cite{subramanian2016robust}. Ahmed et al. presented a sliding-mode control approach~\cite{ahmed20102}. Additionally, Quanser provides teaching material utilizing a linearized model and the \ac{lqr} algorithm to compute control gains for their Aero~2 system.

In another study, Fandel et al. evaluated an approximate dynamic programming approach against \ac{lqr}~\cite{fandel2018development}. Luo et al. investigated the application of multistep Q-learning for optimal control~\cite{luo2017optimal}. Existing research has explored the use of \ac{rl} for the Aero~2 system and similar setups, such as the works by Schäfer et al. \cite{Schaefer24}, Polzounov et al. \cite{polzounov2020blue}, Ouyang et al. \cite{ouyang2017reinforcement}, and Bhourji et al.~\cite{bhourji2024reinforcement}.

Comparative studies of \ac{rl} and traditional control methods, such as \ac{mpc}, have been conducted in various contexts. Lin et al. compared \ac{rl} with \ac{mpc} in terms of performance and computational efficiency for adaptive cruise control~\cite{lin2020comparison}. Ernst et al. \cite{ernst2008reinforcement} compare \ac{mpc} with \ac{rl} for a power system problem. Zhang et al. \cite{zhang2022building} provide a survey on the differences of \ac{mpc} and \ac{rl} for building energy management. Although these studies provide insights into the distinctions of \ac{mpc} and \ac{rl} in different domains, they lack a comprehensive evaluation protocol for motion control.

While substantial work has been done on various control strategies for the Quanser Aero~2 system, there remains a need for a comprehensive comparative analysis that evaluates these methods using a variety of metrics, including computational effort, implementation complexity, dynamic response characteristics, and overall system performance in real-world scenarios.

\subsection{Contributions} \label{sec:contributions}
This paper makes several contributions to the field of applying \ac{drl} to the control of dynamic systems by comparing distinct control methods in a practical experimental setup. Firstly, building upon our previous work \cite{SSRHH24}, we refine the state space representation of the \ac{rl} setting to improve the overall performance of the \ac{ppo} agent. Secondly, we present a detailed comparative analysis of \ac{ppo} and \ac{mpc} in controlling the Quanser Aero~2 system in a 1-\ac{dof} configuration. Additionally, we compare these strategies with \ac{lqr} control, serving as a baseline for our study. This comprehensive evaluation, which includes metrics such as computational effort, implementation complexity, and dynamic response characteristics, provides a systematic overview of each control method's strengths and limitations in real-world scenarios. Therefore, we propose an evaluation protocol that enables a systematic comparison of control strategies, facilitating a deeper understanding of their performance and applicability in practical settings. Finally, based on this comparative analysis and the proposed evaluation protocol, we provide practical recommendations for selecting appropriate control strategies.

\section{Experimental Setup} \label{sec:experimental_setup}
\begin{figure}[ht]
    \begin{subfigure}{0.24\textwidth}
        \centering
        \includegraphics[width=\textwidth]{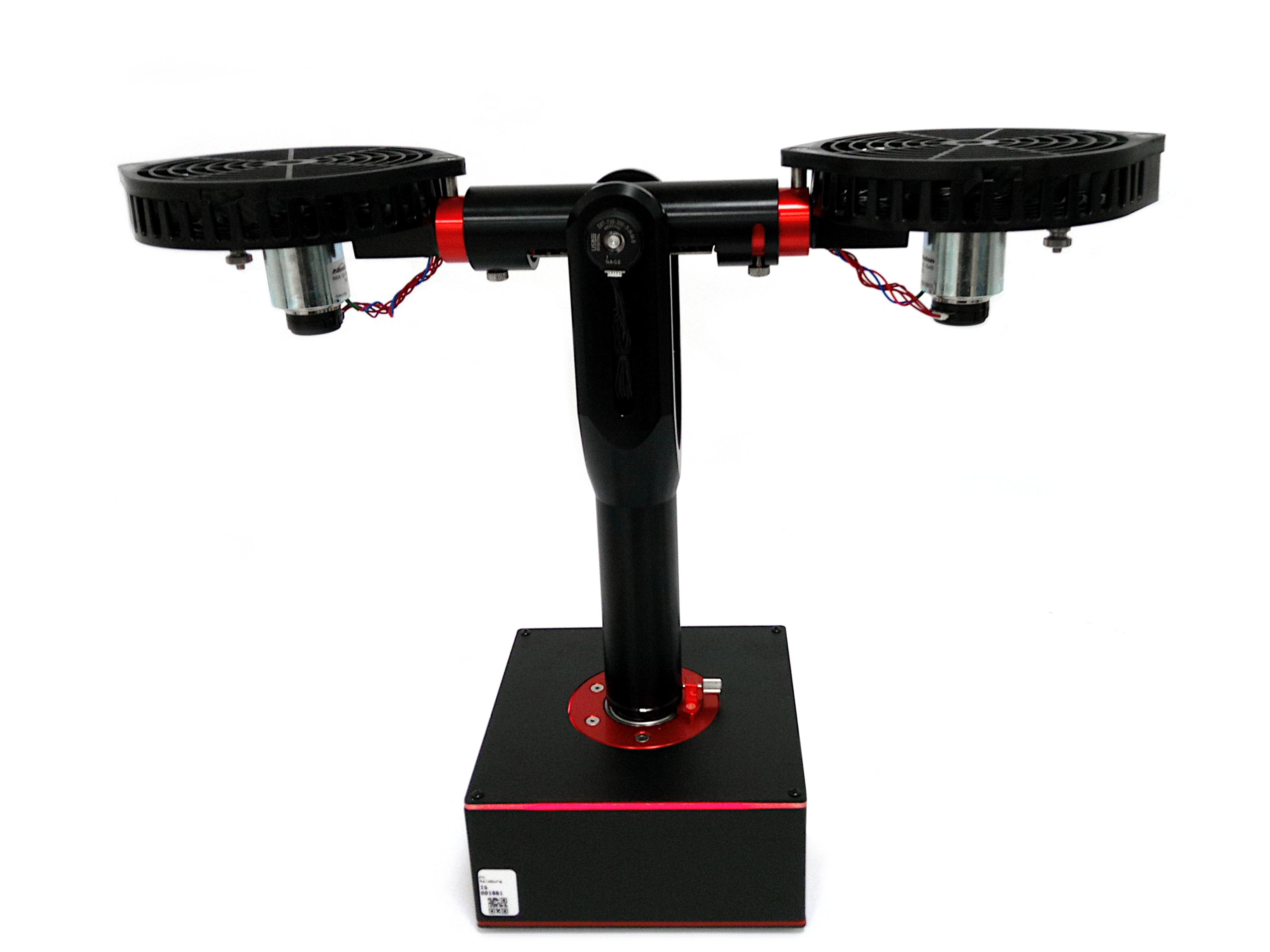}
    \end{subfigure}
    \begin{subfigure}{0.24\textwidth}
        \centering
        \includegraphics[width=\textwidth]{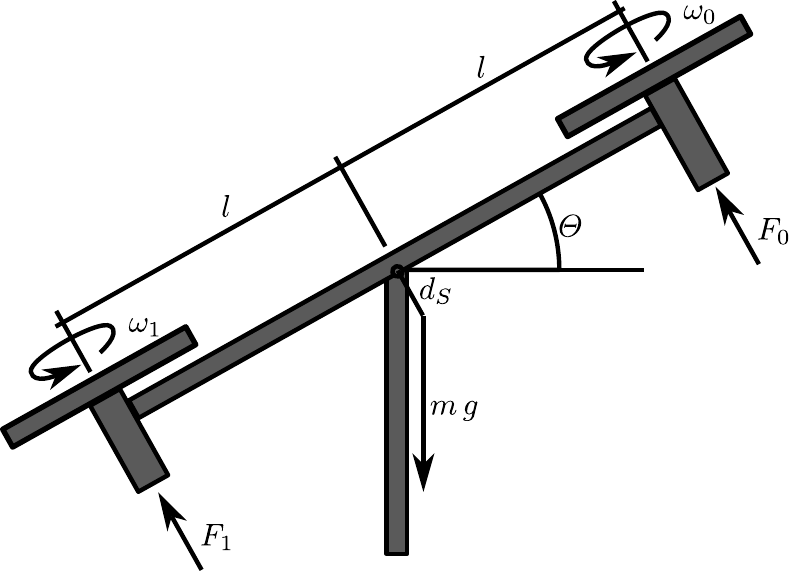}
    \end{subfigure}
    \caption{The Quanser Aero~2~(left) and its schematic representation (right) in a 1-\ac{dof} configuration.}
    \label{fig:aero2}
\end{figure}

The experimental setup utilizes the Quanser Aero~2 test bed in its 1-\ac{dof} configuration, as illustrated in \cref{fig:aero2}. Equipped with two motors driving each fan, the system's inputs are simplified to a single voltage $u$, where $u_0 = u$ is applied to the first motor and $u_1 = -u$ to the second, operating the fans in opposition. Each motor can operate within a voltage range of -24~V to 24~V.

The simulation and the physical system are utilized in our experimentation.
An interface to the environment, which is based on a Simulink simulation, is provided by our previously established approach, while the physical system is accessed through a Python interface. Additionally, direct access to the simulation model and the control of the physical system via Simulink/MATLAB ensures software-stack-independent evaluation, whether using Python or MATLAB.

While the simulation model provides direct access to the actual pitch $\varTheta$ and pitch velocity $\omega = \dot{\varTheta}$, the physical system's $\varTheta$ is measured by a sensor, albeit lacking a sensor for $\omega$. Defining the control problem, the system aims to orient itself to a target pitch $r$. Experiment runs last for 80 seconds, with the target pitch $r$ changing every 10 seconds in the following sequence: 0°, 5°, -5°, 20°, -20°, 40°, -40°, and 0°. Additional evaluation runs analyzing step response characteristics are employed, analyzing the controller's behavior for different jump inputs. Therefore, the system is instructed to orient itself from its origin position ($\varTheta=0$°) to a specific target, including $\pm 5$°, $\pm 10$°, $\pm 20$°, and $\pm 40$° for 60 seconds.

\subsection{System Model}
\begin{figure}[ht]
    \centering
    \includegraphics[width=\columnwidth]{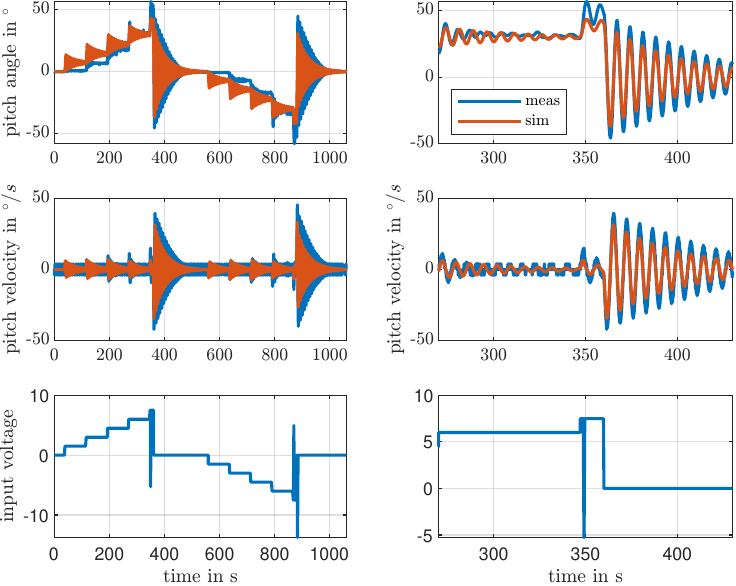}
    \caption{Test sequence used for parameter identification (right column: detailed view of the section showing frequency and damping behavior).}
    \label{fig:model_ident}
\end{figure}

The \ac{lqr} and \ac{mpc} approach both require a mathematical model of the Aero~2. From the principle of angular moment, the following equations of motion can be derived (a second order transfer function of the linearized model is also provided in~\cite{aero2courseware}):
\begin{subequations} 
\label{eq:nonlinear_model}
\begin{align}
    \dot{\varTheta}&=\omega\\
    \dot{\omega}&=\frac{1}{J_p}\left(-k_d\omega-d_S\,m\,g\,\sin \varTheta\right) + 2\,k_u\,u.
\end{align}
\end{subequations}
The angular velocity of the beam is denoted as $\omega$. The distance $d_S$ from the pivot to the beam's center of gravity (see \cref{fig:aero2}) and the mass $m$ of the beam were taken from~\cite{aero2courseware}. The friction coefficient $k_d$, the moment of inertia $J_p$ and the gain $k_u$ translating the DC motor voltage into an angular acceleration were identified from experimental data. The rotor dynamics were neglected due to their faster dynamics compared to the motion of the beam. The linear dependency of the thrust force on the applied voltage is a simplification that turned out to be sufficient for creating a model suitable for controller synthesis. More details on thrust computation can be found in~\cite{Gill2017}. The model parameters were found by minimizing the sum of absolute error between a measured (\enquote{meas}) and a simulated (\enquote{sim}) test sequence shown in \cref{fig:model_ident}, i.e.,
\begin{align}
    \argmin{J_p, k_d, k_u}\sum_k{|\varTheta_{meas,k}-\varTheta_{sim,k}|+|\omega_{meas,k}-\omega_{sim,k}|}.
\end{align}
The left column of \cref{fig:model_ident} shows the entire test sequence. As input voltage, a sequence of steps of amplitude $0, 1.5, 3, 4.5, 6$ and $7.5\,V$ was applied. To avoid hitting the pitch angle limits, a safety mechanism applying a voltage of $5\,V$ of opposite sign was implemented. It activates when the beam is close to and still moving towards the mechanical limit. This safety mechanism is the reason for the voltage spikes at around $350\,s$ and $870\,s$.

From the left column of \cref{fig:model_ident} it is evident, that the assumption on the linear behavior between the applied voltage and the generated thrust by the fan is a strong simplification. For low voltages, the excitation of the real system is significantly smaller than at higher voltages. Since the controller synthesis (see \cref{sec:controller_synthesis}) still performs well, no further model refinements were necessary. The state vector $\mathbf{x}=(\varTheta \,\,\, \omega)^T$ was introduced and \cref{eq:nonlinear_model} was linearized at $u=0$ to obtain the linear state space representation

\begin{subequations}
    \label{eq:statespace}
    \begin{align}
        \dot{\mathbf{x}}&= \underbrace{ \left(\begin{array}{cc}
        0 & 1\\
        -0.8185  & -0.0503
        \end{array}\right) }_{\mathbf{A}} \mathbf{x} + \underbrace{ \left(  \begin{array}{c}
        0\\
        0.0682
        \end{array}\right) }_{\mathbf{b}} u\\
        y&= \underbrace{ (1 \qquad 0) }_{\mathbf{c}^T} \mathbf{x}.
    \end{align}
\end{subequations}
The model~\eqref{eq:statespace} was used for the controller synthesis in \cref{sec:controller_synthesis}.

\subsection{Evaluation Metrics} \label{sec:eval_metrics}
To comprehensively evaluate and compare the performance of different control strategies, we employ a range of metrics that capture different aspects of control quality and computational efficiency:

\begin{itemize}
    \item \textbf{Computational effort:} Measures the time and resources needed to compute the next control action. Lower effort is essential for real-time applications and systems with limited resources.
    \item \textbf{Implementation complexity:} Assesses the effort and expertise needed to design and implement each control strategy, including parameter tuning and overall setup time.
    \item \textbf{Average deviation to target pitch:} Calculates the mean absolute error between the actual pitch angle $\varTheta$ and the desired target pitch $r$ on the 80-second experiment run described in \cref{sec:experimental_setup}, indicating overall control accuracy.
\end{itemize}

To evaluate key dynamic response characteristics, specific metrics to evaluate the response to reference-steps are employed:

\begin{itemize}
    \item \textbf{Steady-state deviation $e_\infty$:} The difference between the output and the target after the system settles, $e_\infty=r-y_\infty$.
    \item \textbf{Overshoot $M_p$:} The extent to which the system exceeds the steady state value before stabilizing, $M_p=(y_{max}-y_\infty)/y_\infty \cdot 100$.
    \item \textbf{Rise-time $t_r$:} The time taken for the response $y$ to rise from 10\% to 90\% of $y_\infty$.
\end{itemize}

To evaluate the response to reference-steps, multiple test runs are conducted for each control strategy. In each run, the target pitch $r$ is initially set to 0° for 10 seconds, followed by a step with a 60-second duration to a constant $r$ value to set one of $\pm 5$°, $\pm 10$°, $\pm 20$°, and $\pm 40$°.

\section{Controller Synthesis} \label{sec:controller_synthesis} % oder Synthesising Controller
In this section, two classic control methodologies (\ac{lqr}, \ac{mpc}) and \ac{ppo} to solve the reference tracking task of the Quanser Aero~2 are explained.

\subsection{Linear Quadratic Regulator}
\label{sec:lqi_design}
\begin{figure}[htbp]
    \centering
    \includegraphics[width=0.4\textwidth]{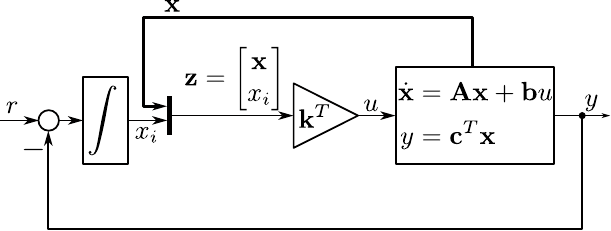}
    \caption{Block diagram of the \ac{lqi} approach.}
    \label{fig:lqi1}
\end{figure}

In our approach, we employed a \ac{lqr} augmented with an integration block, a configuration also known as \ac{lqi}. Based on the methodology in~\cite{YOUNG1972}, this setup ensures zero steady-state error due to the integrator, as illustrated in \cref{fig:lqi1}, the controller provides zero steady-state error. The feedback gain $\mathbf{k}^T$ is obtained by minimizing the cost function~\cite{Mathworks2023}
\begin{align}
    J(u)=\int_{0}^{\infty} {\mathbf{z}^T\mathbf{Q}\mathbf{z}+u\,R\,u}
\end{align}
with
\begin{align}
    \mathbf{Q}=\begin{bmatrix}
        10 &0& 0 \\
        0 &1& 0\\
        0 &0& 100
    \end{bmatrix} \quad \textnormal {and} \quad R=0.001.
\end{align}
The weights in $\mathbf{Q}$ and $R$ were chosen such that the integral of the control error $x_i$ (see \cref{fig:lqi1}) is penalized strong, whereas the actuating signal is weighted less. The state $\mathbf{x}$ that is needed for computing $u$ is obtained from the measurement of the pitch angle $x_1$, and $x_2$ is obtained via numerical differentiation of $x_1$.

\subsection{Model Predictive Control}
\begin{figure}[htbp]
    \centering
    \includegraphics[width=0.38\textwidth]{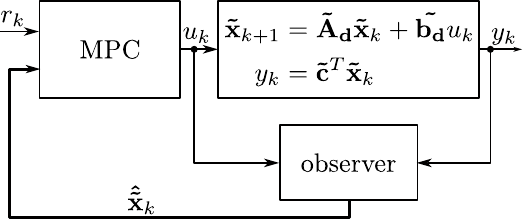}
    \caption{Block diagram of the \ac{mpc} approach.}
    \label{fig:mpc1}
\end{figure}

The block diagram of the \ac{mpc} setup is shown in \cref{fig:mpc1}. The state space model shown in the diagram is given by
\begin{subequations}
    \label{eq:statespaceaugmpc}
    \begin{align}
        {\mathbf{\tilde{x}}}_{k+1}&= \left(\begin{array}{cc}
        \mathbf{A}_d & \mathbf{b}_D\\
        \mathbf{0}  & 1
        \end{array}\right) \mathbf{\tilde{x}}_k + \left(  \begin{array}{c}
        \mathbf{b}_d\\
        0
        \end{array}\right)  u\\
        y_k&= (\mathbf{c}_d^T \qquad c_D)  \mathbf{\tilde{x}}_k,
    \end{align}
\end{subequations}
where $\mathbf{\tilde{x}}_k^T=(\mathbf{x}_k\qquad d_k)^T$ is the state, augmented by a disturbance term $d_k$. $\mathbf{A}_d$, $\mathbf{b}_d$ and $\mathbf{c}_d^T$ are the discrete time parameters of the system model~\eqref{eq:statespace}. The disturbance $d_k$ is assumed to have integrating behavior and to act on the model output. Its purpose is to achieve a zero steady state error when tracking the reference pitch angle $r_k$, see~\cite{Maeder2010} for more details on this approach. Zero steady state offset is accomplished by choosing $\mathbf{b}_D=\mathbf{0}$ and $c_D=1$, estimating $\mathbf{\hat{\tilde{x}}}$ via an observer and taking the estimated $d_k$ into account for the computation of the desired steady state value of the state $\mathbf{x}_{ss}$. A discrete time Luenberger observer~\cite{Franklin1998} of the form 
\begin{align}
    \mathbf{\hat{\tilde{x}}}_{k+1}=\mathbf{\tilde{A}_d}\mathbf{\hat{\tilde{x}}} + \mathbf{\tilde{b}}_d u + \mathbf{k}_L\,(y-\mathbf{\tilde{c}}_d^T \mathbf{\hat{\tilde{x}}}_k)
\end{align}
was used to estimate the augmented state. \emph{Remark:} Since the proposed observer is implemented, the pitch velocity $x_2$ can be obtained from this observer and does not need to be computed from $x_1$ by any other means.

The \ac{mpc} computes the actuating signal in two steps: first, the desired steady state value of the state $\mathbf{x}_{ss}$ and the actuating signal $u_{ss}$ is computed from
\begin{align}
    \begin{bmatrix}
        \mathbf{x}_{ss}\\
        u_{ss}
    \end{bmatrix} = \begin{bmatrix}
        \mathbf{I}-\mathbf{A}_d & -\mathbf{b}_d\\
        \mathbf{c}_d^T & 0
    \end{bmatrix}^{-1}
    \begin{bmatrix}
        \mathbf{0}\\
        r_k-\mathbf{c}_d^T d_k,
    \end{bmatrix}
\end{align} and second, the optimal actuating sequence $(u_{k+i})$ is obtained by solving
\begin{align}
    (u_{k+i})&=\argmin{u_{k+i}} \sum_{i=1}^{N-1} (\mathbf{x}_{k+i} - \mathbf{x}_{ss})^T \mathbf{Q} (\mathbf{x}_{k+1}- \mathbf{x}_{ss}) \nonumber \\ 
    &+ (u_{k+i}-u_{ss}) R (u_{k+i}-u_{ss}) \nonumber\\
    &+ (\mathbf{x}_{k+N}-\mathbf{x}_{ss})^T \mathbf{P} (\mathbf{x}_{k+N}-\mathbf{x}_{ss}) \nonumber\\
    s.t.&\nonumber\\
        {\mathbf{x}}_{k+i+1}&= \mathbf{A}_d \, \mathbf{x}_{k+i} + \mathbf{b}_d\, u_{k+i}\nonumber\\
        -24&\leq u_{k+i} \leq 24 \nonumber.
\end{align}

The first element $u_k$ is then applied to the real system.
The horizon $N=60$ was chosen, and the controller operates at a sampling interval of $T_s=20\,ms$. This allows a prediction of $1.2\,s$ into the future. The weights $\mathbf{Q}$ and $R$ were selected as
\begin{align}
    \mathbf{Q}=\begin{bmatrix}
        10 &0 \\
        0 &1 
    \end{bmatrix} \quad \textnormal {and} \quad R=0.01.
\end{align}
The terminal cost matrix $\mathbf{P}$ was computed by solving the discrete-time Riccati equation using the command \verb|dlqr| of~\cite{Mathworks2023} to achieve performance similar to \ac{lqr}.

\subsection{Proximal Policy Optimization}
\ac{rl} is formulated as a \ac{mdp}: at time $t$, we receive state $s_t$, an agent takes action $a_t$ based on a policy $\pi$, resulting in a new state $s_{t+1}$ and reward $g_t$. \ac{rl} learns a policy that maps states to actions while maximizing the cumulative discounted reward $\sum_{t=0}^{\infty} \gamma^t g_t$, where $\gamma$ is the discount factor~\cite{sutton2018reinforcement}.
The specific \ac{drl} algorithm we employ is \ac{ppo}, chosen for its capability to handle continuous action spaces and ensure stable performance due to theoretical policy improvement guarantees~\cite{schulman2017proximal}. Additionally, studies \cite{jiang2021quadrotor, zhang2024ppo} have demonstrated the feasibility of \ac{ppo} in the domain of motion control.

\ac{ppo} employs an actor-critic architecture. The actor network learns the policy $\pi_{\theta}(a|s)$, representing the probability of selecting action $a$ given state $s$, parameterized by $\theta$. The critic network estimates the value function $V_{\phi}(s)$, approximating the expected cumulative reward from state $s$. \ac{ppo} optimizes the policy by maximizing the expected cumulative reward and minimizing the mean squared error in value estimation. Both the actor and critic neural networks have two hidden layers with 64 units each and $\tanh$ activations~\cite{schulman2017proximal}. The \ac{ppo} implementation from \ac{sb3}~\cite{raffin2021stable} was chosen for training. It's essential to note that no hyperparameter tuning was conducted to keep the agent as simple as possible, allowing for a comparison to other control strategies and serving as a baseline for further \ac{rl}-based research for this specific test bed.

To formulate a \ac{rl} problem, states, actions and the reward have to be defined. In our case, we have a single action $u$, the state comprises the distance to the desired angle ($\Delta_t = \varTheta_t - r_t$), the pitch variation ($\hat{\omega}_t=\varTheta_t - \varTheta_{t-1}$) and the actual pitch $\varTheta_t$. Contrary to the state space used in \cite{SSRHH24} we do not rely on the velocity $\omega$. This enhances the transferability to the real system, since there is no need for velocity estimation. The \ac{rl} agent observes the state and produces a corresponding action every 0.1 seconds.
The reward function is defined as the negative absolute distance to the target angle ($g_t = -|\Delta_t|$), motivating the agent to minimize the deviation from the desired pitch $r$. 
This simplicity in the reward function allows for a fair comparison to other control strategies and contributes to the design of a \ac{rl} baseline for future research.
In this study, we solely focus on the overall performance, and thus, this reward function is used exclusively. However, it does not encompass other metrics, such as overshoot or steady-state deviation. While adapting the reward function to include these metrics is a direction for future work, it is beyond the scope of this study.

Multiple training runs with the same parametrization were conducted in the simulation, each consisting of 1 million training steps. As each training step takes 0.1 seconds and a run takes 80 seconds, this results in 1250 training runs for each agent. We conducted the training on a desktop computer equipped with a Quadro P4000 GPU and an Intel(R) Xeon(R) Silver 4108 CPU. On average, the system achieved 163 steps per second, resulting in a training time of approximately 102 minutes per run. The most proficient agent achieved a cumulative reward of -53.1, corresponding to an average absolute pitch deviation of $\text{abs}(-53.1)/800 \cdot 180/\pi = 3.80$° in the simulation. Test runs on the real system demonstrated, that the agent could solve the problem effectively, even surpassing its simulation performance in terms of minimizing the average distance to the target, achieving an average pitch deviation of 3.75° on the real system. This agent was further trained via transfer learning on the real system, resulting in a cumulative reward improvement to -44.3 (average pitch deviation of 3.2°) after an additional 500.000 training steps. Due to real-time constraints, the training on the real system took 15 hours and 10 minutes. This agent is further analysed in \cref{sec:results_and_discussion}.

\section{Results \& Discussion} \label{sec:results_and_discussion}

\begin{figure*}[ht]
    \centering
    \includegraphics[width=1\textwidth]{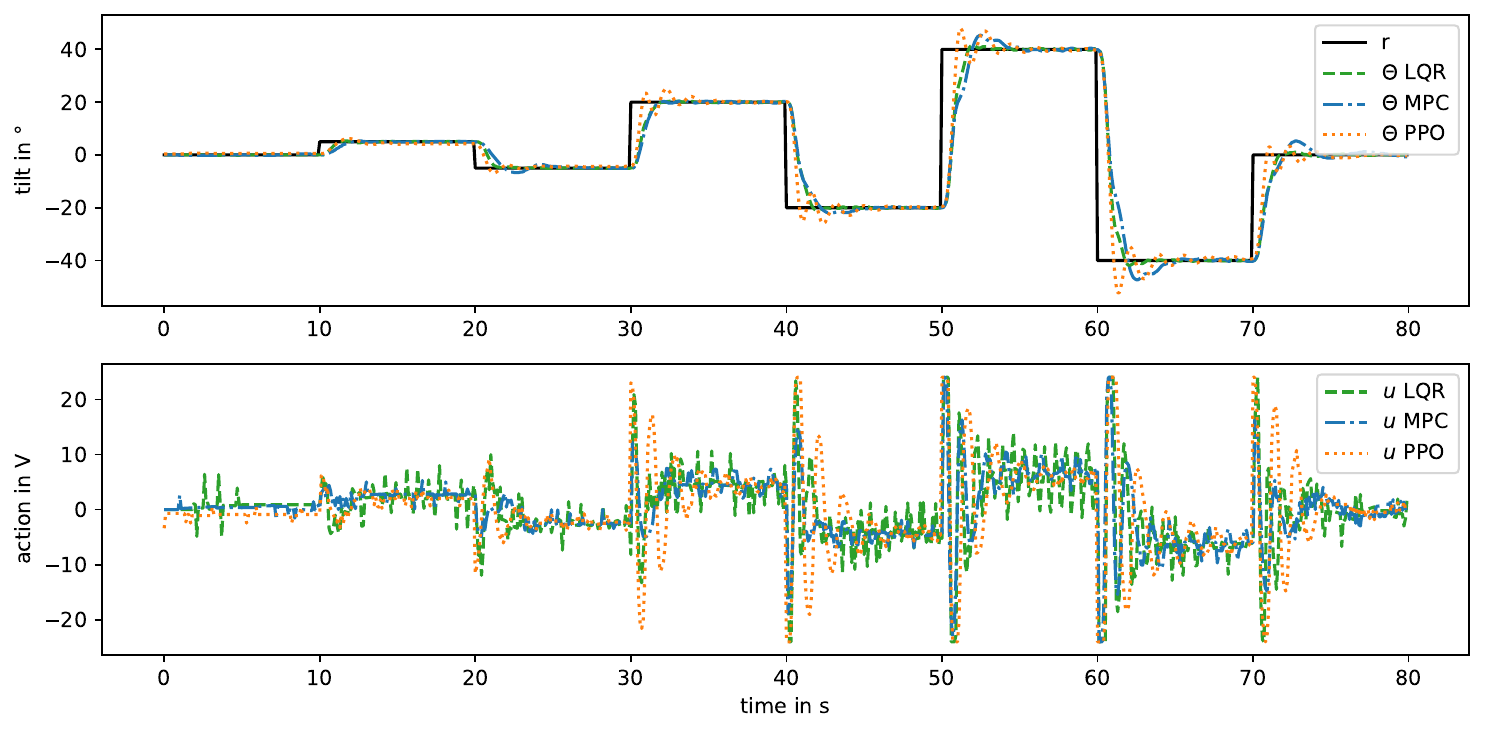}
    \caption{Response of the control strategies (\ac{lqr}, \ac{mpc}, and \ac{ppo}) over an 80-second run with changing target pitches ($r$). The top plot shows the pitch angle as the system tracks the target pitch sequence (0°, 5°, -5°, 20°, -20°, 40°, -40°, 0°). The bottom plot displays the corresponding control action applied by each strategy.}
    \label{fig:testrun}
\end{figure*}

The performance of the control strategies outlined in \cref{sec:controller_synthesis} was evaluated using the metrics described in \cref{sec:eval_metrics}. The results from the experiments are summarized in \cref{tab:comparison}. Two different sets of experiments were conducted: one involving step signals where the controller should track a pitch reference step from 0° to $\mp 5$°, $\pm 10$°, $\pm 20$°, and $\pm 40$° for 60 seconds, and another experiment involving the average absolute distance to the target during a 80-second experiment run with varying target pitches. The second experiment is illustrated in~\cref{fig:testrun}.

All three control strategies achieved acceptable steady-state deviation. The \ac{lqr} controller performed the best, achieving an average absolute error of no more than 0.1°. Both, \ac{mpc} and \ac{lqr} should in theory yield zero steady state error, what is almost achieved by the actual value which is very close to zero. \emph{Remark:} The pitch angle sensor has a resolution of approximately $0.18$°. Two reasons, why the \ac{lqr} performed better than the \ac{mpc} with regard to the steady state error, are: (1) the \ac{lqr} is run at a lower sampling period of $2\,ms$ compared to the \ac{mpc} which uses $20\,ms$, and (2), the selection of the weights $\mathbf{Q}$ and $R$ of the \ac{lqr} put more weight on the control error and less on the actuating signal. However, the \ac{mpc} and \ac{ppo} controllers also performed well, but showed slightly higher deviations compared to \ac{lqr}.

In terms of overshoot, both the \ac{lqr} and \ac{mpc} controllers performed exceptionally well, maintaining minimal overshoot across various target pitches. The \ac{ppo} agent, on the other hand, exhibited significant overshoot, especially at lower target pitches. This behavior indicates that while \ac{ppo} is effective at quickly reaching the target, it tends to overreact to changes in the setpoint, particularly for smaller pitch angles.

The \ac{ppo} agent outperformed both the \ac{lqr} and \ac{mpc} controllers in terms of rise-time, achieving an average rise-time of 0.53 seconds. This demonstrates the \ac{ppo}'s ability to quickly respond to changes in the target pitch, making it suitable for applications requiring fast response times. However, this quick response comes at the cost of increased overshoot, as previously mentioned.

During the 80-second experiment run, with varying target pitches, as shown in \cref{fig:testrun}, the \ac{lqr} controller showed the best performance with an average pitch deviation of 2.91°. The \ac{ppo} controller followed closely with an average pitch deviation of 3.16°, and the \ac{mpc} controller showed an average pitch deviation of 3.72°. The actuating signal of the \ac{ppo} approach shows slightly stronger oscillatory behavior than \ac{lqr} and \ac{mpc}. This can be attributed to the fact that the action is not considered in the reward.

The computational effort required for determining the next control action was found to be low for both the \ac{ppo} and \ac{lqr} strategies. The \ac{ppo} agent, leveraging neural networks, performs computations efficient once trained. The \ac{lqr} controller requires to perform the numerical differentiation of the pitch angle $x_1$ to obtain $x_2$, and the computation of a dot-product of two 3-element vectors. Both operations are computationally lightweight. The \ac{mpc} controller, however, required significant computational resources due to the need for solving the optimization problem~\eqref{fig:mpc1}, which can be a limiting factor in real-time applications.

To evaluate robustness, we introduced disturbances by unbalancing the Quanser Aero~2 system through adjusting the movable masses that are attached to each side of the bar. This intentional unbalancing causes a non-zero pitch angle when no voltage is applied. Both \ac{lqr} and \ac{mpc} demonstrated effective performance under these conditions, maintaining stability with minimal steady-state deviation. In contrast, \ac{ppo} struggled, showing high steady-state deviation and less effective control actions. Improving \ac{ppo}’s robustness could involve additional training on the disturbed system and incorporating the integral of the steady-state error into the state space and reward function. These enhancements are out of scope of this paper and are planned for future work.

\paragraph{Benefits and drawbacks of \ac{lqr}} This approach offers a very simple control law, its design is intuitive (selection of $\mathbf{Q}$ and $R$) and the obtained results are satisfactory. However, constraints on the actuating signal are not inherently covered by the \ac{lqr}, and - in case of nonlinear systems, the choice of the weighting matrices eventually needs to be done in a more conservative manner to avoid oscillating behavior when deviating too much from the operating point used for linearization. This behavior was evident when performing the controller tuning: putting too much weight on the control error led to oscillating actuating signals. The oscillations were more pronounced in regions further away from the operating point. The design of the control law requires the availability of a linear(ized) model of the plant.

\paragraph{Benefits and drawbacks of \ac{mpc}} Like for \ac{lqr}, the controller tuning is intuitive. Furthermore, the implementation of input-, output-, and state-constraints is straightforward. In the present example, only input-constraints were used (see \eqref{fig:mpc1}). These benefits come at the cost of higher computational complexity. In the present example, the sampling time needed to be chosen larger ($20\,ms$ compared to $2\,ms$) in order to solve the optimization problem in time. The execution of the solver on the target hardware needs some additional effort, too. In our case, a suitable toolbox was available to automate this implementation step. Furthermore, in the present setup, an additional component, a state observer, was needed to achieve a zero steady state error. This increases the complexity of the controller parametrization.

\paragraph{Benefits and drawbacks of \ac{ppo}} In contrast to the classic approaches described above, the \ac{ppo} does not make any requirements on a linear system dynamics. Consequently, nonlinear system behavior does not increase the complexity of the steps involved for training the agent. It needs to be ensured that the state contains the information describing the nonlinear-behavior. In the presented example,\, equation~\eqref{eq:nonlinear_model} contains a nonlinear function of $\varTheta$. Consequently, $\varTheta$ should also be part of the state. 
Additionally, if the agent is trained exclusively on the real system, there is no need for a model of the plant. This is a significant advantage for systems that are difficult to describe mathematically.

\begin{table}[]
\caption{Comparison of the three control strategies (\ac{lqr}, \ac{mpc}, and \ac{ppo}) in terms of steady-state deviation $e_\infty$, overshoot $M_p$, rise-time $t_r$, deviation to the target $\Delta$ on the 80-second training signal, computational effort and implementation complexity. Lower numbers indicate a better performance. The best value(s) for each metric are indicated in bold.}
\label{tab:comparison}
\centering
\begin{tabular}{lc|lll}
\textbf{Metric}       & \textbf{r} & \textbf{LQR}     & \textbf{MPC}    & \textbf{PPO}    \\ \hline \hline
\multirow{8}{*}{$e_\infty$} & 5°   & -0.10 °          & \textbf{0.08 °} & 0.78 °          \\ 
                            & 10°  & \textbf{-0.02 °} & -0.20 °         & 0.68 °          \\
                            & 20°  & \textbf{0.14 °}  & -0.39 °         & \textbf{0.14 °} \\
                            & 40°  & \textbf{0.10 °}  & -0.43 °         & 0.45 °          \\
                            & -5°  & \textbf{-0.08 °} & 0.10 °          & -0.61 °         \\
                            & -10° & \textbf{0.20 °}  & \textbf{0.20 °} & -0.51 °         \\
                            & -20° & -0.14 °          & 0.22 °          & \textbf{0.04 °} \\
                            & -40° & \textbf{0.08 °}  & \textbf{0.08 °} & -0.45 °         \\ \hline
Ø $|e_\infty|$               &      & \textbf{0.10 °}  & 0.21 °          & 0.46 °          \\ \hline \hline
\multirow{8}{*}{$M_p$}      & 5°   & \textbf{6.90\%}  & 10.71\%         & 45.83\%         \\
                            & 10°  & 1.76\%           & \textbf{1.73\%} & 30.19\%         \\
                            & 20°  & 1.77\%           & \textbf{0.86\%} & 21.24\%         \\
                            & 40°  & \textbf{2.64\%}  & 4.78\%          & 14.67\%         \\
                            & -5°  & 10.71\%          & \textbf{6.90\%} & 48.00\%         \\
                            & -10° & \textbf{0.00\%}  & 1.73\%          & 29.63\%         \\
                            & -20° & 2.65\%           & \textbf{0.87\%} & 20.18\%         \\
                            & -40° & \textbf{1.75\%}  & 6.58\%          & 17.33\%         \\ \hline
Ø $M_p$                     &      & \textbf{3.52\%}  & 4.27\%          & 28.38\%           \\ \hline \hline
\multirow{8}{*}{$t_r$}      & 5°   & \textbf{0.80 s}  & 1.00 s          & 1.00 s          \\
                            & 10°  & 0.80 s           & 1.30 s          & \textbf{0.50 s} \\
                            & 20°  & 1.10 s           & 1.00 s          & \textbf{0.40 s} \\
                            & 40°  & 1.10 s           & 1.40 s          & \textbf{0.50 s} \\
                            & -5°  & 0.80 s           & 1.20 s          & \textbf{0.50 s} \\
                            & -10° & 0.90 s           & 1.30 s          & \textbf{0.40 s} \\
                            & -20° & 1.00 s           & 1.00 s          & \textbf{0.40 s} \\
                            & -40° & 1.10 s           & 1.40 s          & \textbf{0.50 s} \\ \hline
Ø $t_r$                     &      & 0.95 s           & 1.20 s          & \textbf{0.53 s} \\ \hline \hline
\multicolumn{2}{l|}{Ø $|\Delta|$}                & \textbf{2.91 °}           & 3.72 °          & 3.16 °          \\ \hline
\multicolumn{2}{l|}{Computational effort}      & \textbf{low}             & high           & \textbf{low}             \\ \hline
\multicolumn{2}{l|}{Implementation complexity} & \textbf{low}             & medium            & medium
\end{tabular}
\end{table}

\section{Conclusion \& Future Work}
This study presented a comprehensive evaluation of three control strategies - \ac{lqr}, \ac{mpc}, and \ac{ppo} - applied to the Quanser Aero~2 system in a 1-\ac{dof} configuration. The performance of these controllers was accessed based on a set of key metrics, including steady-state deviation, overshoot, rise-time, average deviation to the target pitch, computational effort, and implementation complexity.

\ac{lqr} demonstrated accuracy and minimal overshoot, but with a longer rise-time. \ac{mpc} performed well with overshoot control and offers the advantage of handling constraints like maximum pitch and velocity, despite higher computational demands. \ac{ppo} excelled in rise-time and adaptability, performing effectively on both simulation and physical system, but exhibited higher overshoot.

\ac{ppo}'s main advantage lies in eliminating the need for a precise model, which could be further enhanced with data-efficient reinforcement learning methods such as probabilistic \ac{mpc} \cite{kamthe2018data}. This would address sample efficiency issues, making \ac{rl} a more viable option for real-world applications. Additionally, refining the \ac{ppo}'s learned policy by adapting the reward function can further improve performance. Future research will investigate the impact of including additional metrics, like the overshoot and action fluctuation, in the reward function.

In summary, \ac{lqr} is recommended for accuracy and minimal steady-state error, \ac{mpc} for scenarios with computational resources allowing for constraint handling, and \ac{ppo} for applications requiring rapid response and adaptability. 
Future work will focus on improving \ac{ppo}'s reward function to reduce overshoot, integrating hybrid approaches to combine the strengths of different control strategies, and addressing the scalability of these methods to multi-\ac{dof} systems. The model development, including the equations and parameter identification, for the 2-\ac{dof} configuration is expected to require slightly more effort than for the 1-\ac{dof} configuration. Given the curse of dimensionality \cite{bellman1957dynamic}, training the \ac{ppo} agent will likely require more computational resources compared to the current setup.

\section*{Acknowledgment}
Financial support for this study was provided by the Christian Doppler Association (JRC ISIA), the corresponding WISS Co-project of Land Salzburg and the European Interreg Österreich-Bayern project BA0100172 AI4GREEN.

\bibliographystyle{IEEEtran}
\bibliography{references, jrcisia-published}

\begin{thebibliography}{10}
\providecommand{\url}[1]{#1}
\csname url@samestyle\endcsname
\providecommand{\newblock}{\relax}
\providecommand{\bibinfo}[2]{#2}
\providecommand{\BIBentrySTDinterwordspacing}{\spaceskip=0pt\relax}
\providecommand{\BIBentryALTinterwordstretchfactor}{4}
\providecommand{\BIBentryALTinterwordspacing}{\spaceskip=\fontdimen2\font plus
\BIBentryALTinterwordstretchfactor\fontdimen3\font minus \fontdimen4\font\relax}
\providecommand{\BIBforeignlanguage}[2]{{%
\expandafter\ifx\csname l@#1\endcsname\relax
\typeout{** WARNING: IEEEtran.bst: No hyphenation pattern has been}%
\typeout{** loaded for the language `#1'. Using the pattern for}%
\typeout{** the default language instead.}%
\else
\language=\csname l@#1\endcsname
\fi
#2}}
\providecommand{\BIBdecl}{\relax}
\BIBdecl

\bibitem{ivanov2018survey}
D.~Ivanov, S.~Sethi, A.~Dolgui, and B.~Sokolov, ``A survey on control theory applications to operational systems, supply chain management, and industry 4.0,'' \emph{Annual Reviews in Control}, vol.~46, pp. 134--147, 2018.

\bibitem{kiumarsi2017optimal}
B.~Kiumarsi, K.~G. Vamvoudakis, H.~Modares, and F.~L. Lewis, ``Optimal and autonomous control using reinforcement learning: A survey,'' \emph{IEEE transactions on neural networks and learning systems}, vol.~29, no.~6, pp. 2042--2062, 2017.

\bibitem{maaruf2022survey}
M.~Maaruf, M.~S. Mahmoud, and A.~Ma’arif, ``A survey of control methods for quadrotor uav,'' \emph{International Journal of Robotics and Control Systems}, vol.~2, no.~4, pp. 652--665, 2022.

\bibitem{gorges2017relations}
D.~G{\"o}rges, ``Relations between model predictive control and reinforcement learning,'' \emph{IFAC-PapersOnLine}, vol.~50, no.~1, pp. 4920--4928, 2017.

\bibitem{Franklin1998}
G.~F. Franklin, J.~D. Powell, and M.~L. Workman, \emph{Digital Control of Dynamic Systems}, 3rd~ed.\hskip 1em plus 0.5em minus 0.4em\relax Ellis-Kagle-Press, 1998.

\bibitem{schulman2017proximal}
J.~Schulman, F.~Wolski, P.~Dhariwal, A.~Radford, and O.~Klimov, ``Proximal policy optimization algorithms,'' \emph{arXiv preprint arXiv:1707.06347}, 2017.

\bibitem{subramanian2016robust}
R.~G. Subramanian and V.~K. Elumalai, ``Robust mrac augmented baseline lqr for tracking control of 2 dof helicopter,'' \emph{Robotics and Autonomous Systems}, vol.~86, pp. 70--77, 2016.

\bibitem{ahmed20102}
Q.~Ahmed, A.~Bhatti, S.~Iqbal, and I.~Kazmi, ``2-sliding mode based robust control for 2-dof helicopter,'' in \emph{2010 11th International Workshop on Variable Structure Systems (VSS)}.\hskip 1em plus 0.5em minus 0.4em\relax IEEE, 2010, pp. 481--486.

\bibitem{fandel2018development}
A.~Fandel, A.~Birge, and S.~Miah, ``Development of reinforcement learning algorithm for 2-dof helicopter model,'' in \emph{2018 IEEE 27th International Symposium on Industrial Electronics (ISIE)}.\hskip 1em plus 0.5em minus 0.4em\relax IEEE, 2018, pp. 553--558.

\bibitem{luo2017optimal}
B.~Luo, H.-N. Wu, and T.~Huang, ``Optimal output regulation for model-free quanser helicopter with multistep q-learning,'' \emph{IEEE Transactions on Industrial Electronics}, vol.~65, no.~6, pp. 4953--4961, 2017.

\bibitem{Schaefer24}
G.~Sch\"{a}fer, J.~Rehrl, S.~Huber, and S.~Hirlaender, ``{Exploring the Dynamics of Reinforcement Learning in Aerospace Control},'' Presented at the RL4AA Salzburg 2024, {Salzburg, Austria}, Feb. 2024.

\bibitem{polzounov2020blue}
K.~Polzounov, R.~Sundar, and L.~Redden, ``Blue river controls: a toolkit for reinforcement learning control systems on hardware,'' \emph{arXiv preprint arXiv:2001.02254}, 2020.

\bibitem{ouyang2017reinforcement}
Y.~Ouyang, W.~He, and X.~Li, ``Reinforcement learning control of a single-link flexible robotic manipulator,'' \emph{IET Control Theory \& Applications}, vol.~11, no.~9, pp. 1426--1433, 2017.

\bibitem{bhourji2024reinforcement}
R.~S. Bhourji, S.~Mozaffari, and S.~Alirezaee, ``Reinforcement learning ddpg--ppo agent-based control system for rotary inverted pendulum,'' \emph{Arabian Journal for Science and Engineering}, vol.~49, no.~2, pp. 1683--1696, 2024.

\bibitem{lin2020comparison}
Y.~Lin, J.~McPhee, and N.~L. Azad, ``Comparison of deep reinforcement learning and model predictive control for adaptive cruise control,'' \emph{IEEE Transactions on Intelligent Vehicles}, vol.~6, no.~2, pp. 221--231, 2020.

\bibitem{ernst2008reinforcement}
D.~Ernst, M.~Glavic, F.~Capitanescu, and L.~Wehenkel, ``Reinforcement learning versus model predictive control: a comparison on a power system problem,'' \emph{IEEE Transactions on Systems, Man, and Cybernetics, Part B (Cybernetics)}, vol.~39, no.~2, pp. 517--529, 2008.

\bibitem{zhang2022building}
H.~Zhang, S.~Seal, D.~Wu, F.~Bouffard, and B.~Boulet, ``Building energy management with reinforcement learning and model predictive control: A survey,'' \emph{IEEE Access}, vol.~10, pp. 27\,853--27\,862, 2022.

\bibitem{SSRHH24}
G.~Sch{\"a}fer, M.~Schirl, J.~Rehrl, S.~Huber, and S.~Hirlaender, ``Python-based reinforcement learning on simulink models,'' in \emph{11th International Conference on Soft Methods in Probability and Statistics (SMPS 2024)}, Salzburg, Austria, Sep. 2024, accepted.

\bibitem{aero2courseware}
\emph{Aero2 - Pitch Parameter Estimation}, Quanser Inc., 2023.

\bibitem{Gill2017}
R.~Gill and R.~D’Andrea, ``Propeller thrust and drag in forward flight,'' in \emph{2017 IEEE Conference on Control Technology and Applications (CCTA)}.\hskip 1em plus 0.5em minus 0.4em\relax IEEE, Aug. 2017.

\bibitem{YOUNG1972}
P.~C. Young and J.~C. Willems, ``An approach to the linear multivariable servomechanism problem†,'' \emph{International Journal of Control}, vol.~15, no.~5, pp. 961--979, May 1972.

\bibitem{Mathworks2023}
\BIBentryALTinterwordspacing
{The Mathworks}, ``Matlab 2023a,'' 2023. [Online]. Available: \url{https://www.mathworks.com}
\BIBentrySTDinterwordspacing

\bibitem{Maeder2010}
U.~Maeder and M.~Morari, ``Offset-free reference tracking with model predictive control,'' \emph{Automatica}, vol.~46, no.~9, pp. 1469--1476, Sep. 2010.

\bibitem{sutton2018reinforcement}
R.~S. Sutton and A.~G. Barto, \emph{Reinforcement learning: An introduction}.\hskip 1em plus 0.5em minus 0.4em\relax MIT press, 2018.

\bibitem{jiang2021quadrotor}
Z.~Jiang and A.~F. Lynch, ``Quadrotor motion control using deep reinforcement learning,'' \emph{Journal of Unmanned Vehicle Systems}, vol.~9, no.~4, pp. 234--251, 2021.

\bibitem{zhang2024ppo}
J.~Zhang, C.~E.~O. Rivera, K.~Tyni, and S.~Nguyen, ``A ppo-based drl auto-tuning nonlinear pid drone controller for robust autonomous flights,'' \emph{arXiv preprint arXiv:2404.00204}, 2024.

\bibitem{raffin2021stable}
A.~Raffin, A.~Hill, A.~Gleave, A.~Kanervisto, M.~Ernestus, and N.~Dormann, ``Stable-baselines3: Reliable reinforcement learning implementations,'' \emph{Journal of Machine Learning Research}, vol.~22, no. 268, pp. 1--8, 2021.

\bibitem{kamthe2018data}
S.~Kamthe and M.~Deisenroth, ``Data-efficient reinforcement learning with probabilistic model predictive control,'' in \emph{International conference on artificial intelligence and statistics}.\hskip 1em plus 0.5em minus 0.4em\relax PMLR, 2018, pp. 1701--1710.

\bibitem{bellman1957dynamic}
R.~E. Bellman, \emph{Dynamic Programming}.\hskip 1em plus 0.5em minus 0.4em\relax Princeton, NJ: Princeton University Press, 1957.

\end{thebibliography}

\end{document}